\documentclass[runningheads]{llncs}
\usepackage{amssymb}
\usepackage{epsfig}
\usepackage{amsmath}
\usepackage{amssymb}
\usepackage{color}
\newcommand{\cc}{{\bf c}}

\newcommand{\FF}{{\mathbb F}}

\newcommand{\C}{{\cal C}}

\newcommand{\lo}{\overline \lambda}
\newcommand{\lm}{v}
\newcommand{\z}{{z}}
\newcommand{\remove}[1]{}

\newcommand{\A}{{\cal  A}}

\newcommand{\tc}{{\tilde c}}

\newcommand{\cG}{{\cal G}}
\newcommand{\ctG}{\tilde{\cal  G}}
\newcommand{\cF}{{\cal F}}

\newcommand{\te}{{\tilde e}}

\newcommand{\G}{{\cal G}}

\newcommand{\tE}{{\tilde E}}

\begin{document}
\authorrunning{A. De Bonis}
\title{Improved bounds for group testing in arbitrary hypergraphs}
\author{\sc Annalisa De Bonis}
\institute{
Dipartimento di Informatica, Universit\`a di Salerno,\\
 Fisciano (SA), Italy \\
\email{adebonis@unisa.it}}
\date{}

\maketitle

\begin{abstract} 
Recent papers initiated the study of a generalization of {\em group testing} where the potentially contaminated sets are 
the members of a given hypergraph $\cF=(V,E)$. This generalization finds  application in contexts where contaminations can be conditioned by some kinds of social and geographical clusterings.
The paper focuses on few-stage group testing algorithms, i.e., slightly adaptive algorithms where tests are performed in stages and all tests performed in the same stage should be decided at the very beginning of the stage.
In particular, the paper presents the first two-stage algorithm that uses $o(d\log|E|)$ tests for general hypergraphs with hyperedges  of size at most $d$,
and  a three-stage algorithm that improves by a $d^{1/6}$ factor on the number of tests of the best known three-stage algorithm. These algorithms are special cases of an $s$-stage algorithm designed for an arbitrary positive integer $s\leq d$.  The design of this algorithm resort to a new non-adaptive algorithm (one-stage algorithm), i.e., an algorithm where all tests must be decided beforehand.
Further, we derive a lower bound for non-adaptive group testing.
For $E$ sufficiently large, the  lower bound is very close to the upper bound on the number of tests of the best  non-adaptive group testing algorithm known in the literature, and it is the first lower bound that improves on the information theoretic lower bound $\Omega(\log |E|).$
\end{abstract}
\section{Introduction}
Group testing is a very well known search problem that consists in detecting the defective members of a set of objects $O$ by performing tests on properly chosen subsets of the given set $O$.  A test yields a ``yes'' response if the tested group contains one or more defective elements, and a ``no'' response otherwise. The goal is to find all defectives by using as few tests as possible. Dorfman \cite{dor} introduced this search paradigm during World War II as a mass blood testing methodology.  Since then, group testing has become a  fundamental problem in computer science and has been applied in a wide variety of situations ranging  from conflict resolution algorithms for multiple-access systems \cite{wolf},  fault diagnosis in optical networks \cite{optical}, quality control in product testing \cite{sobel}, failure  detection in wireless sensor networks \cite{faultSensor}, data compression \cite{hl}, and many others. 
Among the modern applications of group testing,  some of the most important ones are related to the field of  molecular biology, where group testing is especially employed in the design of  screening experiments. Du and Hwang  \cite{dh1}  provide an extensive coverage of the most relevant applications of group testing  in this area. 
In classical  group testing, the set of defectives is any of the possible subsets of size less than or equal to a certain parameter $d$.
In the present paper, we consider a more general version of group testing, parameterized by a hypergraph $\cF=(V,E)$,
 with the contaminated set being one of the hyperedges of $E$. More precisely, in this generalized version of group testing the goal is to detect the hyperedge that consists  of all defective elements. %In other words, the defective hyperedge contains no non-defective element and includes all defective elements.
This generalization finds  application in contexts where contaminations can be conditioned by social relationships, geography, and climate  \cite{arasli}, \cite{niko}.   
In this contexts the hyperedges of $E$  represent potentially infected  sets that might correspond to  groups of friends, families, neighbours,  and so on. 
\smallskip
\subsection{ Related work  and our contribution}
%In classical group testing, the set of the potentially defective hyperedges consists of all hyperedges of size $d$. 
We first review the main results for classical group testing with respect to algorithms with different levels of adaptiveness. It is well known that with respect to {\em adaptive strategies}, i.e. algorithms that at each step decide which group to test by looking at the responses of previous tests, the best adaptive strategies achieve the information theoretic lower bound $\Omega(d\log(n/d))$, whereas, with respect to {\em non-adaptive} strategies, i.e., strategies in which all tests are decided beforehand, the best strategies 
 are much more costly than their adaptive counterparts.
The  minimum number of tests used by the non-adaptive procedures is estimated by  the minimum length of certain combinatorial structures known under the name of {\em separable codes}, or by the minimum length of codes satisfying a slightly stronger property and known under the name of
 $d$-{\em cover free families}, or equivalently, $d$-{\em superimposed codes} and {\em strongly selective families} \cite{cms},  \cite {dyry}, \cite{erff}, \cite{kts}.
The known bounds for these combinatorial structures \cite{alon}, \cite{dyry}, \cite{rus} imply that the number of tests of any non-adaptive group testing algorithm is lower bounded by
$\Omega((d^2/ \log d) \log n)$, and that there exist non-adaptive group testing algorithms that use $O(d^2\log n)$ tests. Interestingly, for classical group testing,  it has been proved  that by allowing just  a little adaptiveness, one can  achieve the same performance of the best completely adaptive algorithms. 
Indeed, the authors of \cite{siam} proved that the information theoretic lower bound can already be attained by  two-stage group testing algorithms, i.e., group testing algorithms in which the tests are performed in two stages each consisting in  a non-adaptive algorithm. 

The study  of group testing in arbitrary hypergraphs    has been initiated only recently in \cite{edge1} and continued in \cite{sofsem,edge2}.
Similar search models were previously considered by the authors of  \cite{niko} who
assumed a known community structure in virtue of which the population is partitioned into separate families and the defective hyperedges are those that contain elements from a certain number of families.
 While in that paper the information on the structure of potentially infected groups is used to improve on the efficiency of the group testing algorithms, other papers, 
  \cite{goenka}, \cite{zrb},
 exploit this knowledge to improve on the efficiency of decoding the tests' responses. 
 
The definition of group testing in arbitrary hypergraphs, as studied in the present paper, has been given by the authors of \cite{edge1} who investigated
 both adaptive and non-adaptive group testing. For the adaptive setting, when hyperedges in $E$ are  
of size exactly $d$, they give an $O(\log{|E|}+d\log^2{d})$ algorithm that is close to the  $\Omega(\log{|E|})$  information theoretic lower bound.
In the non-adaptive setting, they exploit a random coding technique to prove an $O(\frac dp\log{|E|})$ upper bound on the number of tests, 
where $d$ is the maximum size of a hyperedge $e \in E$ and $p$ is a lower bound on the size of the difference $e'\setminus e$ between any two hyperedges  $e,e' \in E$. This upper bound implies an $O(d\log{|E|})$ bound, if no assumption is made on the size of the difference between any pair  of hyperedges.
In \cite{edge2}, the author presents a new adaptive algorithm for generalized group testing, which is
asymptotically optimal if $d = o(\log_2 |E|)$  and, for $d = 2$,  gives an asymptotically
optimal  algorithm that works in three stages. The author of  \cite{sofsem}  formally defines binary codes that are substantially equivalent to non-adaptive algorithms  for group testing in general hypergraphs. These combinatorial structures are a generalized version of  {\em classical} separable codes \cite{dh1} and are parameterized by the set of hyperedges that correspond to the subsets of columns among which the code should enforce separability.
 Paper \cite{sofsem}  introduces also a notion of selectors, parameterized by a set of hyperedges $E$, that generalizes the $(k,m,n)$-selectors of \cite{siam} by enforcing the desired properties only on subsets of codewords associated with the hyperedges in $E$, similarly to what happens with the selectors of  \cite{algow} and \cite{grv}. The combinatorial constructions  in \cite{sofsem} allow to achieve, in the non-adaptive setting, the same $O(\frac dp\log{|E|})$ upper bound of \cite{edge1}, and to design an  $O(\sqrt d \log |E|)$  three-stage algorithm and  a two-stage algorithm that achieves the  $\Omega(\log |E|)$ information theoretic lower bound, provided that, for some constant $q$ and  for any $q+1$  distinct hyperedges $e,e'_1\ldots,e'_q$, it holds that 
 $|\bigcup_{i=1}^q e'_i\setminus e|=\Omega(d)$. 
 
 \smallskip
 The present  paper focuses on few-stage group testing algorithms, i.e., algorithms where tests are performed in few  completely non-adaptive stages.
In particular, the paper presents the first two-stage algorithm that uses $o(d\log|E|)$ tests for arbitrary hypergraphs with hyperedges  of size at most $d$. Differently from the two stage algorithm in \cite{sofsem}, the two-stage algorithm given in the present paper achieves the above said performance without resorting on any assumption on the size of the differences among hyperedges. 
The present paper gives also a three-stage algorithm that improves by a $d^{1/6}$ factor on the number of tests of the  three-stage algorithm in \cite{sofsem}. These algorithms are special cases of an $s$-stage algorithm, designed for an arbitrary positive integer $s\leq d$, whose performance in terms of the number of tests decreases with $s$, thus providing a trade-off between the number of tests and adaptiveness.
The design of this algorithm resorts to a new non-adaptive algorithm that uses $O(\frac bp\log{|E|})$ to discard all hyperedges $e$ that contain at least $p$ non-defective vertices provided that   the size of the difference $e'\setminus e$ between any two hyperedges  $e,e' \in E$ is at most $b$.
As  far as it concerns the problem of identifying the defective hyperedge non-adaptively, the present paper mainly focuses on  non-existential results, for which very little is known in the literature.
The paper presents a lower bound that, for $E$ being sufficiently large and $p\leq \frac 13d$,    is the first lower bound that improves on the  $\Omega(\log |E|)$ lower bound   and gets  close to the best upper bound for non-adaptive group testing, by exhibiting an $ O(\log \frac {d-p} {p})$ gap with the $O(\frac d p\log|E|)$ upper bound. This gap resembles the gap existing between the best upper bound and the best lower bound  for the non-adaptive case in classical group 
testing. We notice that, in the classical setting, $|E|$ consists of all possible subsets of $\{1,\ldots,n\}$, and by setting $p=1$ in the above  gap, one recovers the $\log d$ gap holding for classical group testing.

\section{Notations and terminology}\label{sec:notation}
For any positive integer $m$, we denote by $[m]$ the set of integers $\{1,\ldots,m\}$. 
A hypergraph is
a pair $\cF=(V,E)$, where 
$V$ is a finite set and $E$ is a family of subsets of $V$. The elements of $E$ will be called
hyperedges. If all hyperedges of $E$ have the same size $d$ then the hypergraph is said to be $d$-uniform.

Unless specified differently, the hypergraph specifying the set of potentially contaminated sets is assumed to have $V=[n]$.
We remark that in  our group testing problem, given an input hypergraph $\cF=([n],E)$, every vertex of $[n]$ is contained in at least one
 hyperedge of $E$. If otherwise, one could remove the vertex from the hypergraph without changing the collection of potentially defective hyperedges.
 As a consequence, for a given hypergraph  $\cF=([n],E)$  we need only to specify its  set of hyperedges $E$ to characterize  both the input of the problems and the related combinatorial tools. 
 
Throughout the paper, we will denote by $e$   the base of the natural logarithm $e=2.7182...$, and unless specified differently, all logarithms are in base 2. %Unless specified differently, the number of elements is assumed to be equal to $n$. %This choice
  \section{The group testing model} 
  In group testing for arbitrary hypergraphs, one is given a hypergraph $\cF=(V,E)$ whose hyperedges contain at most a certain number $d$ of vertices. The set $V$ corresponds to a set of elements  among which there are at most $d$ defective elements. The {\em unknown} defective subset, i.e., the subset  of all defective elements, is one of the hyperedges in $E$. The goal is to find the defective hyperedge by testing groups of elements of $V$. The response to a test is ``yes''  if the tested group contains one or more defective elements, and a ``no'' otherwise. 
  This version of the group testing problem corresponds to {\em classical group testing} when $E$ is the set of all possible subsets of $V$ of size at most $d$. The following section illustrates the correspondence between non-adaptive group testing algorithms and binary codes of size $|V|$, or equivalently families of size $|V|$.
  
 \subsection{Non-adaptive  group testing for general hypergraphs}\label{sec:ndgt}
Let $\cF=(V,E)$ be a hypergraph with $V=[n]$ and hyperedges of size at most $d$. We first illustrate the correspondence between   families of $n$ subsets and non-adaptive group testing algorithms. Given a family  $\FF=\{F_1,\ldots,  F_n\}$ with $F_i\subseteq [t]$, we design a non-adaptive group testing strategy as follows.  We denote the elements in the input set by the integers in $[n]=\{1,\ldots, n\}$, and for $i=1,\ldots, t$, we define the group $T_i=\{j\,:\, i\in F_j\}$. Obviously,     $T_1,\ldots,T_t$ can be tested in parallel and therefore  the resulting algorithm is non-adaptive. Conversely, given a non-adaptive group testing strategy for an input set of size $n$ that tests $T_1,\ldots,T_t$, we define a family $\FF=\{F_1,\ldots,F_n\}$  by setting $F_j=\{i\in [t]\,:\, j\in T_i\}$, for $j=1,\ldots, n$. 
Alternatively, any non-adaptive group testing algorithm for an input set of size $n$ that performs $t$ tests can be represented by a binary code of size $n$ with each codeword being a binary vector of length $t$. This is due to the fact that any family of size $n$ on the ground set $[t]$ is associated with the binary code of length $t$ whose codewords are the characteristic vectors of the members of the family. Given such a binary code  $\C=\{\cc_1,\ldots,\cc_n\}$, one has that  $j$ belongs to  pool $T_i$ if and only if the $i$-th entry $\cc_j(i)$ of $\cc_j$ is equal to 1. Such a code can be represented by a $t\times n$ binary matrix $M$ such that 
$M[i,j]=1$ if and only if element $j$ belongs to $T_i$.
 We represent the  responses to tests on $T_1,\ldots,T_t$ by a binary vector whose $i$-th entry is equal to 1 if and only if $T_i$ tests  positive. We call this vector the {\em response vector}. For any input set of hyperedges  $E$ on $[n]$, the response vector is the bitwise $OR$ of  the columns associated with the vertices of the defective hyperedge $e\in E$. It follows that 
  a non-adaptive group testing strategy successfully detects  the defective hyperedge in $E$ if and only if for any two distinct hyperedges $e,e'\in  E$ we obtain two different response vectors. In terms of the associated binary matrix $M$, this means that the bitwise $OR$ of the columns with indices in $e$ and the $OR$ of the columns with indices in $e'$ are distinct. 
 \section{Combinatorial structures for group testing in arbitrary hypergraphs}
The following definition \cite{sofsem} provides a combinatorial tool which is essentially equivalent to a non-adaptive group testing algorithm in our search model.
\begin{definition}\label{def:separ}\cite{sofsem}
Given a hypergraph $\cF=(V,E)$ with $V=[n]$ with hyperedges of size at most $d$, we say 
that a $t\times n$ matrix $M$ with entries in $\{0,1\}$ is an  $E$-separable code if for any two distinct hyperedges $e,e'\in E$,  
it holds that $\bigvee_{j\in e} c_j\neq \bigvee_{j\in e'} c_j$, where $c_j$ denotes the column of $M$ with index $j$.
The integer $t$ is the length of the $E$-separable code.
 \end{definition}
Having in mind the correspondence between binary codes and non-adaptive algorithms illustrated in Section \ref{sec:ndgt}, one can see that a non-adaptive algorithm successfully determines the contaminated hyperedge in $E$ if and only if the binary code associated to the algorithm is $E$-separable.
 Therefore, the minimum number of tests of such an algorithm coincides with the minimum length of an $E$-separable code. In Section \ref{sec:lower}, we present a lower bound on the minimum length of $E$-separable codes for the case when all hyperedges of $E$ contain exactly $d$ vertices. This lower bound translates into a lower bound on the minimum number of tests needed to find the defective hyperedge in a non-adaptive fashion.
 
\section{Lower Bound on the Number of Rows of $E$-Separable Codes }\label{sec:lower}
Let $E$ be a set of hyperedges of size $d$ on the set of vertices $[n]$ and
let $M$ be a $t\times n$  $E$-separable code with set of columns  $\{c_1,\ldots,c_n\}$.
We define $\cG=\{G_1,\ldots,G_n\}$ as the family whose members are the subsets of $[t]$ having
 the columns of $M$ as characteristic vectors, i.e., $G_j=\{i:\, c_j(i)=1\}$.  Notice that since $M$ is an $E$-separable code, for any two edges $e$, $e'$, it holds that
 $\bigcup_{j\in e'} G_j\neq  \bigcup_{j\in e} G_j$. We say that $\cG$ is an $E$-separable family.
 
 The proof of the lower bound on the length of $E$-separable codes uses a reduction technique that consists in reducing the size of the members of the family by subtracting up to a certain number
 of ``large'' members of the family from the remaining members. While in analogous proofs of non-existential results for classical cover free families  \cite{alon,rus} , one is free to choose the members to be subtracted among 
 all  ``large'' members of the family, here we need the ``large'' members to be associated to vertices of large degree so as to ensure that the compressed
 members form a separable family for a hypergraph of suitable size.
  We need the following  lemma.
 \begin{lemma}\label{lemma:int}
 Let $E$ be a set of hyperedges of size $d$ on $[n]$, and let $f$ be a positive integer with $f\leq |E|d/n$. The number of
 hyperedges of $|E|$ consisting only of vertices with degree larger than or equal to $f$ is at least $|E|-(f-1) n_s\geq |E|-(f-1) (n-1)$, where $n_s$ is the number of vertices in $[n]$ with degree smaller
 than $f$.
 \end{lemma}
 \begin{proof}
 Let $E_g$ denote  the subset of hyperedges of $E$ consisting only of vertices with degree larger than or equal to $f$ and let $E_{mix}$ be the subset of hyperedges
 of $E$ that contain at least one vertex with degree smaller than $f$. 
 Since hyperedges in $E_{mix}$ contain at least one vertex with degree smaller than $f$, 
 the number of the hyperedges in $E_{mix}$ is smaller than or equal to the sum of the degrees of these vertices. Then, having
 denoted with $n_s$  the number of vertices of $E$ with degree smaller than $f$, it holds that $|E_{mix}|\leq (f-1)n_s$. %This is due to the fact that 
 %any hyperedge of $E_{mix}$ contains at least a vertex with degree smaller than $f$.
 Since $|E|=|E_g|+|E_{mix}|$, we have that $|E_{g}|\geq |E|- (f-1)n_s$. Notice that among the vertices in $E$, there is at least a vertex of maximum degree
 $\Delta$.
 Since the sum of  the degrees of all vertices in $[n]$ is equal to  $|E|d$ and it is smaller than or equal to $n\Delta$,  the maximum degree $\Delta$ is at least $|E|d/n$ and consequently, for $f\leq|E|d/n$,  it holds that $n_s<n$, from which it follows that $|E_{g}|\geq E|-(f-1) (n-1)$.
 \end{proof}
 \begin{theorem}\label{thm:lowb}
Let $\cF=([n],E)$ be a $d$-uniform hypergraph, and let $\lm$ be  a positive integer, with $\lm<d \leq \frac{3\lm}{2c}$, and $c$ being a constant in $[1,3/2)$. 
It holds that  
the minimum length of an $E$-separable code is at least
$$
 \frac {\lm }{c(d-\lm/c)\log \left(\frac {e\lm} {c(d-\lm/c)} \right)}  \log\left(
\frac{|E|}{n^{\lm/c}}\right).$$
\end{theorem}
\begin{proof}
Let $\cG$ be a  maximum size  $E$-separable family on the ground set $[t]$. 

We will show how to reduce the size of the members of $\cG$ while preserving separability with respect to a hypergraph with suitable parameters.
Let $z=\lm /c$, for some  constant $c>1$.
 For $q=0,\ldots, z$, we define the set of hyperedges $\tE^q$ as follows
 \begin{equation}\label{eq:E_j}
 E^q=
\begin{cases}
E&\mbox{if $ q=0$}\\
 \{e\setminus \{i_{q-1}\}:\, i_{q-1}\in e, e\in \tE^{q-1}\} 
 ,&\mbox{if $1\leq q \leq z$,}
\end{cases}
\end{equation}
where for each $q=0,\ldots, z$, $i_{q}$ is
a vertex that occurs in at least  $\frac{|E^{q}|}{n-q}$  hyperedges of $E^{q}$.
For $q\geq 1$, the hyperedges in $E^{q}$  have size $d-q$ and  it holds that $|E^{q}|\geq \frac{|E^{q-1}|}{(n-q+1)}$,
from which we get that
\begin{equation}\label{eq:ejsize}
|E^{q}|\geq  \frac{|E|}{n (n-1) \cdots (n-q+1)}.
\end{equation} 
Moreover, we denote with $E^q_g$ the subset of hyperedges of $E^q$ that contain only vertices with degree at least $\frac {|E^q|}{n-q}$.
By Lemma  \ref{lemma:int}, the size of $E^q_g$ is
\begin{equation}\label{eq:size}
|E^q_g| \geq |E^q| -\frac{|E^{q}|}{n-q}(n-q-1)+(n-q-1)> \frac{|E^{q}|}{n-q}.
\end{equation}  
In addition to sets $E^1,\ldots, E^z$, we define families $\cG^0,\ldots, \cG^{z}$ by setting $\cG^0=\cG$ and  $\G^{q}=\{G\setminus  G_{i_{q-1}}:\, G\in \cG^q\setminus \{G_{i_{q-1}}\} \}$, for $q=1,\ldots,z$, where $G_{i_{q-1}}$ is the member of $\cG^q$ associated with the vertex $i_{q-1}$ in Definition \ref{eq:E_j}. 
For each $q$, $0\leq q \leq z$, one can see that  $\cG^{q}$ is $E^q$-separable by using induction. 
For $q=0$,  $\cG^{q}$ is obviously $E^q$-separable since $E^0=E$ and $\cG^0=\cG$.
Suppose by induction hypothesis that $\cG^{q}$ is  $E^q$-separable up to a certain $q\geq 0$. In order to see that  $\cG^{q+1}$ is $E^{q+1}$-separable, assume by contradiction that $\cG^{q+1}$ is not  $E^{q+1}$-separable. By this assumption, there exists
two hyperedges $e'_1,e'_2\in E^q$ such that $\bigcup_{k\in e'_1} G'_k=\bigcup_{k\in e'_2} G'_k$, where $G'_k$ is the member of $\cG^{q+1}$ associated to vertex $k$. 
It follows that $\bigcup_{k\in e'_1} G'_k\cup G_{i_{q}}=\bigcup_{k\in e'_2} G'_k\cup G_{i_{q}}$. By construction of $E^{q+1}$ and $\cG^{q+1}$, both the 
hyperedges $e_1=e'_1\cup \{i_{q}\}$ and $e_2=e'_2\cup \{i_{q}\}$ belong to $E^{q}$
and  it holds that $\bigcup_{k\in e'_1} G'_k\cup G_{i_{q}}=\bigcup_{k\in e_1} G_k$ and $\bigcup_{k\in e'_2} G'_q\cup G_{i_{q}}=\bigcup_{k\in e_2} G_q$, 
where $G_k$ is the member of $\cG^{q}$ associated to $k$. Therefore, it holds that $\bigcup_{k\in e_1} G_q=\bigcup_{k\in e_2} G_q$, and this contradicts the induction hypothesis that   $\cG^q$ be $E^q$-separable.

Suppose that $\cG$ contains at least one member of size larger than $ t/z$. If this is the case, we start a process aimed at bounding from above the cardinality
 of $\cG$
by  that of a family with members of size smaller than or equal to $t/z$. In the following  discussion, we assume that $z$ divides $t$ and that $c$ divides $\lm $. One can easily convince herself that the proof works also in the case when $t$ is not a multiple of $z$ and $\lm $ is not a multiple of $c$.
We start by setting $\cG^0=\cG$ and define families with  members of progressively smaller  size as follows. For $q\geq 0$, we replace $E^q$ by $E^{q+1}$, if there exists a vertex $i_q$ with degree at least $\frac{|E^{q}|}{n-q}$ and  $|G_{i_q}|>t/z$, where $G_{i_q}$ is the member of $\cG^q$ associated with the vertex $i_q$. Notice that $i_q$ is as required by  (\ref{eq:E_j}). 
We have seen that for each $q$, $0\leq q \leq z$,  $\cG^{q}$ is $E^q$-separable.  
If for some $q\geq 0$, $\cG^q$  contains no member of size larger than $t/z$ associated with a vertex $x$ of degree at least $\frac{|E^{q}|}{n-q}$ in $E^q$,
then the compression process stops and, if there exist members of $E^q$ of size larger than $t/z$, then the set $E^q$ is replaced by $E^q_g$ i.e.,  by the subset of the hyperedges of $E^q$ consisting only of vertices with degree at least
 $\frac{|E^{q}|}{n-q}$. In this case, we replace $\cG^q$ by the family $\cG^q_g$ obtained from $\cG^q$ by removing the members associated to vertices with degree smaller than  $\frac{|E^{q}|}{n-q}$. The family $\cG^q_g$  is obviously $E_g^q$-separable. By (\ref{eq:size}), we have that  $|E^q_g|>\frac{|E^q|}{n-q}$. 

Now, we need to show that, at the end of the compression process, we are left with a set of hyperedges that is not ``too small''.
 Notice that after at most $q=\z$ steps, the compression process terminates since, at each step,  at least $t/\z$ elements
are removed from the ground set of $\G$, and therefore, after at most $z$ steps,  one is left with a family $G_q$ whose members have size at most $t/z$. 
However, it might happen that the  process terminates   at some step $q<z$ because $\cG^q$ contains no member of size larger than $t/z$ that is associated with a vertex $x$ occurring in at least $\frac{|E^{q}|}{n-q}$ hyperedges of $E^q$. 
 We have seen that, in this case, if $\cG^q$  contains  some member of size larger than $t/z$,   then $E^q$ is replaced by $E^q_g$ and the family $\cG^q$ is replaced by $\cG_g^q$, i.e.,
by the family obtained by taking only the members of $\cG^q$ associated with vertices of degree at least $|E^q|/(n-q)$ and all of them have size at most $t/\z$. Let us denote by $\tE$ the set of hyperedges resulting from the above process and let $\tilde G$ denote a maximum size  $\tE$-separable family.

Suppose that the compression process stops after $j< \z$ steps. We further reduce the number of vertices in the hyperedges by defining
 \begin{equation}\label{eq:tE_j}
 \tE^q=
\begin{cases}
\tE&\mbox{if $ q=j$}\\
 \{e\setminus \{i^*_{q-1}\}:\, i^*_{q-1}\in e, e\in \tE^{q-1}\} 
 ,&\mbox{if $j+1\leq q \leq z$,}
\end{cases}
\end{equation}
where, for each $q=j,\ldots, z-1$, $i^*_{q}$ is
a vertex with largest degree in $E^q$.
Notice that in the above definition we do not need  $i^*_{q}$ to be associated to a large member of $\cG^q$.
Moreover, for $q=j,\ldots,z-1$, we  set $\ctG^{q+1}=\{G\setminus  G_{i^*_q}:\, G\in \ctG^q\setminus \{G_{i^*_q}\} \}$, where $G_{i^*_q}$ is the member of $\ctG^q$ associated with the vertex $i^*_q$.  By the same argument we have used to show that $\G_q$ is $E^q$-separable, one can see that $\ctG^q$ is  an $\tE^q$ separable family,

In order to limit  the size of $\tE^{z}$ from below, we assume that the compression process stops with $j$ as large as possible, i.e., $j=z$. In this case,
at the end of compression process, the algorithm is left with $\tE=E^{\z}$ and  it holds that 
\begin{equation}\label{eq:lowbej}
|E^{\z}|\geq\frac{|E|}{n(n-1)\ldots (n-\z+1)}.
\end{equation}
Since, all members of $\cG^{z}$ have size at most $t/\z=  ct/\lm $,  the unions $\bigcup_{k\in e} G_k$ have size at most 
$(d-\lm /c) ct/\lm $, where $G_k$ denotes the member of $\cG^{z}$ associated with vertex $k$.
Further, since $\cG^{z} $ is $E^{z}$-separable,  we have that the following inequality holds:
 $$
 |E^{z}|\leq \sum_{i=0}^{ \lfloor(d-\lm /c) ct/\lm \rfloor}{t\choose i}.
$$
 By hypothesis, it is $d< \frac{3\lm }{2c}$, and consequently, it holds that $ (d-\lm /c)  ct/\lm \leq t/2$, and we can exploit the following well known inequality, holding for $a\leq t/2$,
%\begin{equation}\label{eq:leftLowAdap1}
$$\sum_{i=0}^a{t\choose i}\leq 2^{tH(a/t)},$$
%\end{equation}
where $H(\frac at)$ denotes the binary entropy $H(\frac at)=-\frac at \log \frac at-(1-\frac at)\log(1-\frac at)$.
 Therefore, we obtain the following upper bound:
 \begin{equation}\label{eq:bound_union}
|E^{z} |\leq \sum_{i=0}^{\lfloor(d-\lm /c) ct/\lm \rfloor}{t\choose i}\leq 2^{tH\left(\frac {c(d-\lm /c)}{\lm } \right)}.
 \end{equation}
For $a/t\leq1/2$, it holds the following inequality (see (4) in \cite{jco} for a proof of it):
%  \begin{equation}\label{eq:H}
  $H\left(\frac{a}{t}\right)\leq\frac at \log\frac {et}a.$
 % \end{equation} 
This inequality implies that
\begin{equation}\label{eq:boundonsum}
 H\left(\frac {c(d-\lm /c)  }{\lm } \right)\leq \left(\frac {c(d-\lm /c)  }{\lm } \right)  \log \left(\frac {e\lm } {c(d-
 \lm /c)}     \right).
\end{equation}
From (\ref{eq:bound_union}) and (\ref{eq:boundonsum}) 
it follows that $|E^{z }|\leq  2^{t \left(\frac{c(d-\lm /c)} {\lm }    \right)  \log \left(\frac {e\lm } {c(d-
 \lm /c)}      \right)} $, 
 and consequently,
one has that
\begin{equation}\label{eq:t1}
t\geq  \left(\frac {\lm } {c(d-\lm /c)}   \right)  \frac {\log |E^{\z }|}{\log \left(\frac {e\lm } {c(d-
 \lm /c) }\right)}.
\end{equation}
Finally, by (\ref{eq:lowbej}) and (\ref{eq:t1}), it holds  that
\begin{eqnarray}\label{eq:fin}
&t&\geq  \left(\frac {\lm } {c(d-\lm /c)}   \right)  \frac {\log |E^{\lm /c }|}{\log \left(\frac {e\lm } {c(d-
 \lm /c)} \right)}\cr\cr
&\geq& \frac {\lm }{c(d-\lm /c)\log \left(\frac {e\lm } {c(d-\lm /c)} \right)} 
\log\left(\frac{|E|}{n(n-1)\cdots (n-\lm /c+1)} \right)\cr\cr
&\geq&\frac {\lm }{c(d-\lm /c)\log \left(\frac {e\lm } {c(d-\lm /c)} \right)}  \log\left(
\frac{|E|}{n^{\lm /c}}\right).
\end{eqnarray}
\end{proof}
%In the above theorem we have assumed $d\leq \frac{3\lo}2$ because this ensures that the constant $c$ is not ``too small'', namely it is larger than or equal to 1. We remark that for $d> \frac{3\lo}2$, one has that $p=d-\lo>\frac d3$, and consequently, the $O(\frac dp\log|E|)$ upper bound of \cite{archedge1,edge1} matches  the information theoretic lower bound. 
Notice that if in the statement of Theorem \ref{thm:lowb},  $E$  is such that, for any two distinct hyperedges $e,e'\in E$, it holds that $|e\cap e'|\leq \lo=\lm$, then, we have that the following corollary of Theorem \ref{thm:lowb} holds.
\begin{corollary}\label{cor:lowb}
Let $\cF=([n],E)$ be a $d$-uniform hypergraph, and let $\lo$ be  a positive integer, with $\lo<d \leq \frac{3\lo}{2c}$ and $c$ being a constant in $[1,3/2)$. 
If  for any two distinct hyperedges $e,e'\in E$,  it holds that $|e\cap e'|\leq \lo$, then 
the minimum length of an $E$-separable code is at least
$$
 \frac {\lo }{c(d-\lo/c)\log \left(\frac {e\lo} {c(d-\lo/c)} \right)}  \log\left(
\frac{|E|}{n^{\lo/c}}\right).$$
 %where $c$ is a  constant in $[1,\frac{3\lambda}{2d}]$.
\end{corollary}
Notice that,  if a $d$-uniform hypergraph has $n$ vertices and its hyperedges  pairwise intersect in at most $\lo$ vertices, then it contains at most $\frac{{n\choose \lo+1}}{{d\choose \lo+1}}$ hyperedges (see Lemma 2.4.1 in \cite{dh1}). Therefore, the set of hyperedges $E$ in the statement of  Corollary \ref{cor:lowb} has size at most  $\frac{{n\choose \lo+1}}{{d\choose \lo+1}}$, and  the term 
$\log\left(\frac{|E|}{n^{\lo/c}}\right)$ in the lower bound of that corollary might be very small.
However, if  the set of hyperedges $E$ has size 
$|E|>n^{\frac{\lo c'}{c}}$, 
for some positive constant $c'\in  (1,c)$, then it holds $ \frac {|E|}{n^{\lo/c }}\geq |E|^{\frac{c'-1}{c'}}$, and 
%consequently, one has that  
%the lower bound  of Theorem \ref{thm:lowb} is at least
%$$
% \frac {\lo }{c(d-\lo/c)\log \left(\frac {e\lo} {c(d-\lo/c)} \right)} \left(\frac{c'-1}{c'} \right)\log|E|\geq
%\frac {\lo }{(d-\lo)\log \left(\frac {e\lo} {(d-\lo/)} \right)} \left(\frac{c'-1}{c'} \right)\log|E|,$$
%with the above inequality following from 
%by taking into account that that the constant $c$ falls in $ [1,\frac{3}{2}]$, we
 by setting $\tc=\frac{c'}c$, we have that the upper bound of  Corollary \ref{cor:lowb}, along with information theoretic lower bound,  implies the following:
 \begin{corollary}\label{cor:low}
Let $\cF=([n],E)$ be a $d$-uniform hypergraph, and let $\lo$ be  a positive integer, with $\lo<d \leq \frac{3\lo}{2c}$ and $c$ being a constant in $[1,3/2)$. 
If  for any two distinct hyperedges $e,e'\in E$,  it holds that $|e\cap e'|\leq \lo$ and 
$|E| \geq n^{ \lo\tc}$, for some constant $\tc\in (\frac1c,1)$, then 
the minimum length of an $E$-separable code is at least
$
\Omega\left(\max\left\{\frac {\lo }{(d-\lo)\log \left(\frac {e\lo} {d-\lo} \right)}\log |E|,\log |E|\right\}\right).$
\end{corollary}
%\begin{proof}
%The lower bound follows from discussion preceding the corollary and from the information theoretic lower bound.
%\end{proof}
Obviously, the lower bounds stated by  Corollary \ref{cor:lowb} and Corollary \ref{cor:low}, translate into lower bounds on the minimum number of tests needed by any non-adaptive algorithm that finds the defective hyperedge in a set of hyperedges $E$ such that  $\max\{|e'\cap e|:\, e,e'\in E\}\leq \lo$, and $|e|=d$ for any $e\in E$. 
Papers \cite{sofsem,edge1} give non-adaptive algorithms that determine the defective hyperedge by  $O(\frac dp\log|E|)$ tests, when the set $E$ is such that $\min\{|e'\setminus e|:\, e,e'\in E\}\geq p$. If the hyperedges in such a set $E$ have size equal to $d$,  then the hyperedges pairwise intersect in at most $\lo=d-p$ vertices, i.e., the quantity $d-\lo$ in our lower bounds is equal to the parameter $p$ in the above mentioned upper bound of \cite{sofsem,edge1}.
We remark that  when $p=d-\lo=\Theta(d)$, the algorithms in \cite{sofsem,edge1}  achieve the information theoretic lower bound, and consequently, they also achieve the lower bound of Corollary \ref{cor:low} since, for $d-\lo=\Theta(d)$, that bound is equal to the information theoretic lower bound.
On the other hand, if $p=d-\lo=o(d)$, then the ratio between the  $O(\frac dp\log|E|)$ upper bound in \cite{sofsem,edge1} and  the lower bound stated by Corollary~\ref{cor:low} is  
$\frac{d}{\lo}\log \left(\frac {e\lo} {d-\lo}\right)\leq \frac{3}{2}\log \left(\frac {e\lo} {d-\lo}\right)$. Notice that if  $E$ consists of all possible subsets of $d$ vertices in $[n]$, with $n\geq d^d$, then $\lo=d-1\geq \frac 23 d$ and $|E|={n\choose d}\geq n^{d-1}$, so that the lower bound of  Corollary~\ref{cor:low} holds. In this case, the gap between the asymptotic upper and lower bounds for the non-adaptive case is $\log d$, which is the same gap existing between the best upper and lower bounds for non adaptive algorithms in classical group testing.

We point out that the authors of \cite{archedge1,edge1} give an example of a hypergraph for which 
the minimum number of  tests  needed to find the defective hyperedge is $\Omega \left(\frac d{\log d}\log |E|\right)$. However that result does not imply a lower bound for arbitrary hypergraphs.

\subsection{An improved non-adaptive group testing algorithm for general hypergraph}\label{sec:nonadaptive}
In this section, we present a new non-adaptive algorithm that   discards all hyperedges $e$ that contain at least $p$ non-defective vertices provided that   the size of the difference $e'\setminus e$ between any two hyperedges $e$ and $e'$  is at most $b$.
In order to design this algorithm, we need the following result of \cite{sofsem}.
\begin{theorem}\label{thm:gen_non_adaptgt}\cite{sofsem}
Let $\cF=(V,E)$  be a hypergraph with $V=[n]$ with all hyperedges in $E$ of size {\em at most} $d$. For any positive integer $p\leq d$, there exists a  non-adaptive algorithm that allows to discard all but those hyperedges $e$ such that $|e\setminus e^*|< p$, where $e^*$ is the defective hyperedge, and uses
$t=O\left(\frac d p\log |E|\right)$ tests.
\end{theorem}
Notice that if it holds $|e\setminus e'|\geq p$, for any two distinct hyperedges in $E$, then the algorithm of Theorem \ref{thm:gen_non_adaptgt} allows to detect the defective hyperedge. 
Since, for any set $E$,  it holds that $|e\setminus e'|\geq 1$, for any two distinct hyperedges in $E$ such that $e\not\subset e'$, one has that by setting $p=1$ in the algorithm of Theorem \ref{thm:gen_non_adaptgt}, one obtains the $O(d\log |E|)$ upper bound in \cite{sofsem,edge1} on the minimum number of tests needed to determine the defective hyperedge in a non-adaptive fashion. Notice that if $E$ contains hyperedges that are properly contained in other hyperedges, then the algorithm might end up with one or more hyperedges in addition to the defective one. This is possible only if the additional hyperedges are properly contained in the defective hyperedge $e^*$. Since we assume that the defective hyperedge is the one that contains all defective vertices, the algorithm returns the largest of the above said hyperedges.

In the following theorem, we present a new non-adaptive algorithm that can be used to reduce the number of hyperedges that are candidate to be the defective hyperedge.
\begin{theorem}\label{thm:new_nonadapt}
Let $d$ and $n$ be integers with $1\leq d\leq n$, and let $E$ be  a set of hyperedges of size {\em at most} 
$d$ on $[n]$. Moreover, let $b$ be  an integer such that  
$1\leq \max \{ |e'\setminus e|:\,e,e'\in E\}<b$.
For any positive integer $p\leq b-1$, there exists a non-adaptive  group testing algorithm that  allows to discard all but those hyperedges $e$ such that $|e\setminus e^*|< p$, where $e^*$ is the defective hyperedge, and uses
%$$t\leq \frac {e(d+p)}{p} \ln 
%\left( \min\left\{e|E|(|E|-1), e^{2(d+p)-1}\left({n+d-1\over d+
$t=O\left(  \frac bp\log |E|+d\right)$
tests. Moreover, the algorithm returns a set of hyperedges $\hat E$ such that any hyperedge of $\hat E$ has size at most $b-1$ and is a subset of a hyperedge of $E$ that has not be discarded by the algorithm. For any hyperedge ${\hat e}\in \hat E$, it holds that  $|{\hat e}\setminus {\hat e}^*|<p$, 
where  $\hat e^*$ is the unique hyperedge of $\hat E$ such that  $\hat e^*\subseteq e^*$.
\end{theorem}
\begin{proof}
The algorithm chooses a  hyperedge $\te$ of maximum size. Since it holds that  $|\te\setminus e^*|<  b$, at least $|\te|-b+1$ of the vertices of $\te$ are defective. 
The algorithm replaces every  hyperedge $e$ by $e\setminus \te$. Let $\hat E$ denote the set of these hyperedges after removing the duplicates.  From the hypothesis that $\max \{ |e'\setminus e|:\,e,e'\in E\}<b$, it follows that any hyperedge in $\hat E$ has size at most $b-1$.
For each  hyperedge $e\in E$, let us denote with $\hat e$ the corresponding hyperedge of $\hat E$, i.e., $\hat e=e\setminus \te$.  Notice that  a hyperedge $\hat e\in \hat E$  might correspond to more than one hyperedge in $E$.
Theorem \ref{thm:gen_non_adaptgt} implies that there exists an  $O(\frac bp \log |{\hat E|})$ 
non-adaptive algorithm $\cal {\hat A}$ that discards all but the hyperedges $\hat e\in \hat E$ such that $|\hat e\setminus {\hat e^*}|<  p$, where ${\hat e^*}$ is the hyperedge in $\hat E$ corresponding to the original defective hyperedge $e^*$, i.e., ${\hat e}^*={e^*}\setminus  \te$.
Notice that there might be non-defective hyperedges $e$ such that $e\setminus \te=e^*\setminus \te={\hat e^*}$. %We will see how to deal with this fact at the end of the proof,
Let $\hat M$ be the binary matrix associated with $\hat A$ and let $M'$ be the matrix obtained by adding $|\te|$ distinct rows at the bottom of $\hat M$, with each of these rows having an entry equal to 1 in correspondence of a distinct vertex of $\te$ and all the other entries equal to 0.
Let us consider the non-adaptive algorithm associated to $M'$.
 The tests  associated to the rows of $\hat M$ allow to discard all non-defective hyperedges of $\hat E$ but those for which $|{\hat e}\setminus {\hat e}^*|<p$, i.e., the non-defective hyperedges are not discarded if they contain less than $p$ non-defective vertices. Notice that for any hyperedge $e$ that originally contained at least $p$ non-defective vertices, one has that either the corresponding hyperedge  ${\hat e}\in{\hat E}$ contains at least $p$ non-defective  
 vertices, or it holds that $e\cap \te$ contains one or more non-defective vertices. In the former case $\hat e$ is discarded by $\hat A$, whereas in the latter case, ${\hat e}$ is discarded after inspecting the responses to the tests associated with the last $|\te|$ rows of $M'$. Indeed, these responses uncover the non-defective vertices in $\te$ and the algorithm discards $\hat e$ if it contains one or more of these non-defective vertices. Notice that if the problem admits that $E$ contains hyperedges that are properly contained in $e^*$, then it holds that ${\hat e}\neq {\hat e}^*$, i.e, ${\hat e}$ is properly contained in ${\hat e}^*$. Otherwise if it were ${\hat e}={\hat e}^*$, there should be a defective vertex in $\te$ that does not belong to $e$. The algorithm gets rid of such a hyperedge  by discarding all hyperedges in $\hat E$ that do not contain all the  defective vertices in $\te$.   
 Those vertices, if any, are uncovered once again by  the responses to the tests associated with the last $|\te|$ rows of $M'$. 
  \end{proof}
  The  above theorem  asymptotically improves on the upper bound of Theorem \ref{thm:gen_non_adaptgt}  when it holds that $ \max \{ |e'\setminus e|:\,e,e'\in E\}< b$ for $b=o(d)$. 
 \remove{ Theorems \ref{thm:gen_non_adaptgt} and \ref{thm:new_nonadapt} imply the following corollary. 
 \begin{corollary}\label{cor:non-adaptivegt}
Let $d$ and $n$ be integers with $1\leq d\leq n$, and let $E$ be  a set of hyperedges of size {\em at most} 
$d$ on $[n]$. Moreover, let $p\geq 1$ and $b>1$  be  integers such that  
$1\leq p\leq min\{ |e'\setminus e|:\,e,e'\in E\}$ and $ \max \{ |e'\setminus e|:\,e,e'\in E\}< b$.
There exists a non-adaptive group testing algorithm that finds the defective hyperedge in $E$  and uses at most 
%$$\frac {e(d+p)}{p} \ln(e|E|(|E|-1))
$t=O\left(\min\{ \frac dp,b\}\log |E|\right)$
tests. %The associated decoding algorithm requires $O(nt)$ time.
\end{corollary}
\begin{proof}
Theorem \ref{thm:gen_non_adaptgt}  implies the $O(\frac dp\log |E|)$ bound, whereas Theorem \ref{thm:new_nonadapt} implies the existence of an $O(b\log |E|)$ non-adaptive algorithm that returns 
\end{proof}}
In the next section, we resort to Theorem \ref{thm:new_nonadapt} and Theorem \ref{thm:gen_non_adaptgt} to design an $s$-stage algorithm, for an arbitrary $s\leq d$.
\section{An $s$-stage group testing algorithms for general hypergraphs}
In this section we present an $s$-stage algorithm, for an arbitrary $s\geq 1$, i.e, an algorithm consisting in $s$ non-adaptive stages. For $s\geq 3$, the algorithm  improves on the three-stage algorithm given in \cite{sofsem} and for $s=2$ it achieves the same asymptotic performance of the three-stage algorithm in \cite{sofsem}.
 \begin{theorem}\label{thm:s-stagegt}
Let $\cF=(V,E)$  be a hypergraph with $V=[n]$  with hyperedges of size {\em at most} $d$. Let $s$ be any positive integer smaller that or equal to $d$, and let $b_1,\ldots,b_s$ be $s$ integers with $d>b_1>\ldots>b_s=1$. There exists an  $s$-stage algorithm that finds the defective hyperedge in $E$ and uses $O(\frac d{b_1} \log |E|+\sum_{i=2}^{s}(\frac {b_{i-1}}{b_{i}}) \log |E|+b_{i-2}))$ tests. 
\end{theorem}
\begin{proof}
Let us describe the  $s$ stage algorithm achieving the stated upper bound. Let $b_0=d$ and let $b_1,\ldots,b_s$ be $s$ integers with $b_0>b_1>\ldots>b_s=1$. In the following discussion  $e^*$ denotes the defective hyperedge in $E$. The first  stage aims at restricting the search to  hyperedges to a set of  vertices  $E_1\subseteq E$ such that $\max \{ |e'\setminus e|:\,e,e'\in E\}<b_1$.
From Theorem \ref{thm:gen_non_adaptgt}, one has that there is an  $O(\frac d{b_1 } \log |E|)$ 
non-adaptive algorithm ${\cal A}_1$ that discards all but the hyperedges $e$ such that $|e\setminus e^*|<  b_1$, where $e^*$ is the defective hyperedge.  
Let stage 1 execute algorithm $A_1$. 
%If all hyperedges that have not been discarded by $\A_1$ have size smaller than $b_1$, then the algorithm proceeds to  stage 2.  
After running algorithm $\A_1$, the algorithm gets rid  of any not yet discarded hyperedge $e$ for which there exists another non-discarded hyperedge $e'$ such that 
$|e'\setminus e|\geq b_1$. Indeed, $e$ cannot be the defective hyperedge. To see this, observe  that if $e$ were the defective hyperedge then $e'$ would have been discarded after running algorithm $\A_1$. After getting rid of these hyperedges, the algorithm is left only with a set of hyperedges $E_1\subset E$ such that for any two distinct hyperedges $e$ and $e'$ it holds that $|e'\setminus e|< b_1$ and $|e\setminus e'|< b_1$. 
%By Corollary \ref{cor:new_nonadapt}, there exists an $O(\frac{d^{\frac{s-1}{s}}{d^{\frac{s-2}{s}}\log |E'|)$ non-adaptive algorithm $\cal A'$ that finds the defective hyperedge $e^*$. 
Stage 2 aims at restricting the search to a set  $E_2$ of hyperedges  each of which contains  at most $b_1-1$ vertices and such that  $\max \{ |e'\setminus e|:\,e,e'\in E_2\}<b_2$.
Notice that $E_1$ satisfies the hypothesis of Theorem \ref{thm:new_nonadapt} with $b=b_1$. By that theorem, there exists an 
$O(\frac{b_1}{b_2}\log |E_1|+d)$ non-adaptive algorithm $\A_2$ that discards all but the hyperedges $e\in E_1$ such that $|e\setminus e^*|<  b_2$. Let stage 2 run algorithm $\A_2$. By Theorem \ref{thm:new_nonadapt}, after running algorithm $\A_2$, stage 2 is left with a set $\tE_2$ of hyperedges of size at most $b_1-1$  such that each hyperedge $\hat e$ of $\tE_2$ is subset of a hyperedge of $E_1$ and such that   $|{\hat e}\setminus {\hat e}^*|<b_2$, where  $\hat e^*$ is the unique hyperedge of $\tE_2$ for which it holds  that $\hat e^*$ consists only of  defective vertices. Stage 2, discards any hyperedge $\hat e$ of $\tE_2$ for which there exists another non-discarded hyperedge $\hat e'$ such that 
$|\hat e'\setminus \hat e|\geq b_1$. Indeed, by the same argument used above, one can see that $\hat e$ cannot be $\hat e^*$.   After discarding these hyperedges from $\tE_2$, stage 2 is left with a set $E_2$ of  hyperedges  of size at most $b_1-1$  that contains a unique hyperedge $\hat e^*$ that is entirely contained in $e^*$. Moreover, for any two hyperedges $\hat e,{\hat e'}\in E_2$, it holds that  $|{\hat e}\setminus {\hat e'}|<b_2$.
By continuing in this way, we design an algorithm that in stage $i$, for any $i\geq 2$,  performs $O(\frac{b_{i-1}}{b_i}\log |E_{i-1}|+b_{i-2})$ tests, and returns a set $E_{i}$ of hyperedges each of which has size  at most $b_{i-1}-1$ and is a subset of a hyperedge of $E_{i-1}$.  
Moreover, for any two distinct  hyperedges ${\hat e},{\hat e'}\in E_{i}$, it holds that  $|{\hat e}\setminus {\hat e'}|<b_{i}$,  and there exist a unique 
hyperedge  $\hat e^*$ of $E_i$ such that  $\hat e^*\subseteq e^*$. We prove by induction on $i$ that this invariant holds at any stage $i\geq 2$.
We have seen that this invariant is true for $i=2$. Let us prove that it  is true for any $i>2$. Suppose that the invariant is true up to a certain stage $i\geq 2$ and let us prove that it holds also for stage $i+1$. Since by induction hypothesis stage $i$ satisfies the invariant, one can see that stage $i$ returns a set $E_i$ of hyperedges each of which has size  at most $b_{i-1}-1$ and
such that, for any hyperedges ${\hat e},{\hat e'}\in E_{i}$, it holds that  $|{\hat e}\setminus {\hat e'}|<b_{i}$. Moreover, $E_{i}$ contains a unique hyperedge 
$\hat e^*$ such that  $\hat e^*\subseteq e^*$. The set $E_i$  satisfies the hypothesis of Theorem \ref{thm:new_nonadapt} with $d=b_{i-1}-1$ and $b=b_i$. That theorem implies that there exists an $O(\frac{b_{i}}{b_{i+1}}\log |E_i|+b_{i-1})$ non-adaptive algorithm $\A_{i+1}$ that  returns a set of hyperedges $\tE_{i+1}$ such that any hyperedge of $E_{i+1}$ has size at most $b_i-1$, and  for any hyperedge it holds that $|\hat e\setminus  \hat e^*|\leq b_{i+1}$, where 
 $\hat e^*$  is the unique hyperedge in $\tE_{i+1}$ such that  $\hat e^*\subseteq e^*$. After running algorithm  $\A_{i+1}$ , stage $i+1$ discards any hyperedge $\hat e$ of $\tE_{i+1}$ for which there exists another  hyperedge $\hat e'$ of  $\tE_{i+1}$ such that 
$|\hat e'\setminus \hat e|\geq b_{i+1}$. Indeed, by the same argument used for stage 1, one can see that $\hat e$ cannot be $\hat e^*$.  After discarding these hyperedges from $\tE_{i+1}$, stage $i+1$ is left with a set $E_{i+1}$ of  hyperedges  of size at most $b_i-1$  that contains a unique hyperedge $\hat e^*$ that is entirely contained in $e^*$. Moreover, for any two hyperedges $\hat e,{\hat e'}\in E_2$, it holds that  $|{\hat e}\setminus {\hat e'}|<b_{i+1}$. Notice also that stage $i+1$ performs $O(\frac{b_{i}}{b_{i+1}}\log |E_i|+b_{i-1})$ tests, Therefore, we have proved that  the invariant holds also for $i+1$
.
At stage $s$, the algorithm surely determines the defective hyperedge since we have set $b_s=1$. Indeed, this means that at stage $s$, the algorithm of Theorem \ref{thm:new_nonadapt} discards all  hyperedges of $E_{s}$ but those such that $|\hat e \setminus \hat e^*|<1$ and returns a set of hyperedges  $E_{s}$ such that any hyperedge of $E_{s}$ has size at most $b_{s-1}-1$ and is a subset of a hyperedge of $E_{s-1}$. Moreover,  for any hyperedge ${\hat e}\in E_{s}$, it holds that  $|{\hat e}\setminus {\hat e}^*|<b_{s}=1$, 
where  $\hat e^*$ is the  unique hyperedge of $E_{s}$ such that $\hat e^*\cup e^*$. In other words, $E_{s}$ contains only the hyperedge $\hat e^*$. In order to recover $e^*$ the algorithm needs only to add back to $\hat e^*$ the vertices that have been removed from $e^*$ through the $s$ stages.

By summing up the number of tests performed by the $s$ stages, we get that the total number of tests performed by the algorithm is 
$O(\frac d{b_1} \log |E|+\sum_{i=2}^{s}(\frac {b_{i-1}}{b_{i}} \log |E_{i-1}|+b_{i-2})$.
\end{proof}
By setting $b_i=d^{\frac{s-i}{s}}$, for $i=1,\ldots,s$, we have that $b_1,\ldots, b_s$ satisfies the hypothesis of Theorem \ref{thm:s-stagegt}, and we get the following corollary.
\begin{corollary}\label{cor:s-stagegt}
Let $\cF=(V,E)$  be a hypergraph with $V=[n]$  with hyperedges of size {\em at most} $d$. Let $s$ be any positive integer smaller than or equal to $d$. There exists an  $s$-stage algorithm that finds the defective hyperedge in $E$ and uses 
$O(s d^{\frac 1s} \log |E|+sd)$
tests.
\end{corollary}

Notice that by  setting $s=\lceil \log d\rceil$, the above corollary implies that there exists a $\lceil \log d\rceil$-stage algorithm that finds the defective hyperedge in $E$ and uses $O((\log d)( \log |E|)+d)$ tests.

By setting $s=2$ in Corollary \ref{cor:s-stagegt}, we obtain a two stage algorithm that achieves the same asymptotic number of tests of the three-stage algorithm of \cite{sofsem}. Interestingly, this existential result is independent from the size of the differences between hyperedges. This result is stated in the following corollary. 
\begin{corollary}\label{cor:2-stagegtsqrt}
Let $\cF=(V,E)$  be a hypergraph with $V=[n]$  with hyperedges of size {\em at most} $d$. There exists a  two-stage algorithm that finds the defective hyperedge in $E$ and uses $O\big({\sqrt d}\log |E|+d)$ tests.
\end{corollary}
By setting $s=3$ in Corollary \ref{cor:s-stagegt}, we obtain a three stage algorithm that improves by a $d^{1/6}$ factor on the upper  bound on the number of tests of the three-stage algorithm in \cite{sofsem}.

\begin{corollary}\label{cor:3-stagegtsqrt}
Let $\cF=(V,E)$  be a hypergraph with $V=[n]$  with hyperedges of size {\em at most} $d$. There exists a  three-stage algorithm that finds the defective hyperedge in $E$ and uses  $O(d^{\frac{1}{3}} \log |E|+d)
$ tests.
\end{corollary}
We end this section by comparing the two-stage algorithm given in \cite{sofsem} with the one in Corollary \ref{cor:2-stagegtsqrt}.
\begin{theorem}\label{thm:two-stageold} \cite{sofsem}
Let $\cF=(V,E)$  be a hypergraph with $V=[n]$  and all hyperedges in $E$ of size at most $d$. Moreover, let $q$ and $\chi$ be positive integers 
such that    $1\leq q\leq |E|-1$, $\chi =\min\{ |\bigcup_{i=1}^q e'_i\setminus e|,  \mbox{ for any $q+1$ distinct $e,e'_1,\ldots,e'_q\in E$}\}$. 
There exists a  (trivial) two-stage algorithm that uses  
$
t< \frac {2e(d+\chi)}{\chi}  \left(1+ \ln\left({d+\chi-1\choose d+\chi-d-1}\beta\right)\right)+dq$ tests,
where $\beta=\min\left\{e^q|E|\left({|E|-1\over q}\right)^q, e^{d+\chi-1}\left({n+d-1\over d+\chi-1}\right)^{d+\chi}\right\}$.
\end{theorem}
We observe that, in order for the algorithm in Theorem \ref{thm:two-stageold} to outperform the two-stage algorithm of Corollary \ref{cor:2-stagegtsqrt}, there should exist a constant $q$
such that  $$\min\{ |\bigcup_{i=1}^q e'_i\setminus e|,  \mbox{ for any $q+1$ distinct $e,e'_1,\ldots,e'_q\in E$}\}\geq \sqrt d.$$ 
We remark that the upper bounds proved for our $s$-stage algorithms do not rely on  any particular feature of the hypergraph.
\remove{On the other hand, we  remark that the two-stage algorithm in \cite{sofsem} is a trivial two-stage group testing algorithm. This means that  the algorithm in the second stage performs only tests on individual vertices. These algorithms are very important from a practical perspective  since, in many applications, the cost of the second stage is not considered an additional cost, in view of the fact that tests on individual elements are needed in any case to confirm that the selected elements are really defective.}
 

\begin{thebibliography}{5}

\message{References}
\bibitem{alon} Alon, N.,  Asodi, V.: Learning a hidden subgraph.   SIAM J. Discrete Math. 18, no. 4, pp. 697--712  (2005)

\bibitem{arasli} Arasli, B. and Ulukus, S.: Graph and cluster formation based group testing, 2021 IEEE ISIT, pp. 1236-1241 (2021)




\bibitem{cms}  Clementi, A. E. F., Monti, A.,  Silvestri, R.: Selective families, superimposed codes, and
broadcasting on unknown radio networks. In:  Twelfth Annual ACM-SIAM Symposium on Discrete Algorithms, pp. 709--718 (2001)

\bibitem{aghp} Coja-Oghlan, Amin and Gebhard, Oliver and Hahn-Klimroth, Max and Loick, Philipp  
  Coja-Oghlan, A., Gebhard, Ol., Hahn-Klimroth, M., Loick, P.: Optimal Group Testing. In: 
  Thirty Third Conference on Learning Theory, 
pp. 1374--1388 (2020)


\bibitem{algow} De Bonis, A.:  Conflict Resolution in Arbitrary Hypergraphs. In: 19th International Symposium, ALGOWIN 2023, Lecture Notes in Computer Science 14061, Springer,  pp. 13--36 (2023)
\bibitem{jco}  De Bonis, A.: Constraining the number of positive responses in adaptive, non-adaptive, and two-stage group testing. Journal of 
 Combinatorial Optimization,  32, 4 pp 1254–1287 (2016).
\bibitem{sofsem} De Bonis, A.: Group Testing in Arbitrary Hypergraphs and Related Combinatorial Structures. In: Fernau, H., Gaspers, S., Klasing, R. (eds) SOFSEM 2024: Theory and Practice of Computer Science. SOFSEM 2024. Lecture Notes in Computer Science, vol 14519. Springer, Cham. pp 154--168 (2024)


\bibitem{siam} De Bonis, A,  G\c asieniec, L, Vaccaro, U.: Optimal two-stage algorithms for group testing problems.   SIAM J. Comput.   34, no. 5, pp.1253--1270 (2005)

\bibitem{dor} Dorfman, R.: The detection of defective members of large populations.   Ann. Math. Statist. 14, pp. 436--440 (1943)
\bibitem{dh1} Du, D.Z.,  Hwang, F. K.:   Pooling design and Nonadaptive Group Testing. Series on Appl. Math. vol. 18. World Scientific (2006)

\bibitem{dyry}  D'yachkov, A.G., Rykov, V.V.: A survey of superimposed code theory.   Problems Control Inform. Theory 12, pp. 229--242  (1983)
\bibitem{erff} Erd\"os, P., Frankl, P.,  F\"uredi, Z.: Families of finite sets in which no set is covered by
the union of $r$ others.   Israel J. Math. 51, pp. 79--89  (1985)
%\bibitem{ekr} Erd\"os, P., Ko, C.,  RADO,R.: Intersection theorems for systems of finite sets. Quart. J. Math,
%Oxford (2) 12 ), 313--320 (1961)
\bibitem{grv} Gargano, L., Rescigno, A.A., Vaccaro, U.: On k-Strong Conflict–Free Multicoloring. In: Gao, X., Du, H., Han, M. (eds) Combinatorial Optimization and Applications. COCOA 2017. Lecture Notes in Computer Science(), vol 10628. Springer, Cham., pp. 276–290 (2017)
\bibitem{archedge1} Gonen, M., Langberg, M., Sprintson A.: Group Testing on General
Set-Systems. Manuscript, Available at https://arxiv.org/abs/2202.04988
\bibitem{edge1} Gonen, M., Langberg, M., Sprintson A.: Group testing on general set-systems.  In: 2022 IEEE International Symposium on Information, pp. 874--879 (2022)
\bibitem{optical} Harvey,  N.J.A., Patrascu, M.,   Wen, Y.,  Yekhanin, S., Chan, V.W.S.: Non-Adaptive Fault Diagnosis for All-Optical Networks via Combinatorial Group Testing on Graphs.  In:  26th IEEE Int. Conf. on Comp. Communications,
 pp. 697--705 (2007)
 

\bibitem{hl}   Hong, E.S., Ladner, R.E.: Group testing for image compression.    IEEE Transactions on Image Processing  11, no. 8,  pp. 901--911 (2002) 
 
%\bibitem{hs} Hwang, F.K., S\'os, V.T.: Non adaptive hypergeometric group testing.   Studia Sc. Math. Hungarica
%22,  pp. 257--263 (1987)

%\bibitem{hwang} Hwang, F.K.: A method for detecting all defective members in a population by group testing. Journal of the American Statistical Association 67, no. 339, pp. 605--608 (1972) 
%\bibitem{johnson1} Johnson, S.M.: A new upper bound for error-correcting codes, IRE Trans. Inf. Theory  IT-8, pp. 203--207 (1962) 
 

\bibitem{goenka}  Goenka R., Cao S.J.,  Wong C.W.,   Rajwade A., Baron D.: Contact
tracing enhances the efficiency of covid-19 group testing. In: ICASSP 2021 %2021 IEEE International Conference on Acoustics, Speech and Signal Processing (ICASSP), pages 
pp. 8168--8172 (2021)
%\bibitem{karbasi} Karbasi, A., Zadimoghaddam, M. Sequential group testing with graph constraints. In 2012 IEEE Information Theory Workshop,  292--296, 2012.
%\bibitem{luo} Luo, S., Y. Matsuura, Y., Miao, Y.,  Shigeno M. Non-adaptive group testing on graphs with connectivity. Journal of Combinatorial Optimization, 38(1):278--291, 2019.
\bibitem{kts} Kautz, W.H., Singleton, R.C.: Nonrandom binary superimposed codes.    IEEE Trans Inf. Theory 10, pp. 363--377 (1964)


\bibitem{faultSensor} Lo, C., Liu, M., Lynch, J.P., Gilbert, A.C.: Efficient Sensor Fault Detection Using Combinatorial Group Testing. In:  2013 IEEE International Conference on Distributed Computing in Sensor Systems, pp. 199--206 (2013)

\bibitem{niko} Nikolopoulos, P.,S. Srinivasavaradhan, R., Guo T., Fragouli, C., Diggavi S.: Group testing for connected communities. In: The 24th Int. Conf. on Artificial Intelligence and Statistics.
volume 130, pp. 2341–2349. PMLR (2021)

\bibitem{rus} Ruszink\'{o}, M.: On the upper bound of the size of the $r$-cover-free families.   J.
Combin. Theory Ser. A 66, pp. 302--310 (1994)

\bibitem{sobel} Sobel M., Groll, P.A.: Group testing to eliminate efficiently all defectives in a binomial sample.   Bell System Tech. J. 38, pp. 1179--1252  (1959)


\bibitem{edge2} Vorobyev, I.:
Note on generalized group testing. Available at 
https://doi.org/10.48550/arXiv.2211.04264 (2022)

%\bibitem{wilson}  Wilson, R.M.: The exact bound in the Erd\"os-Ko-Rado theorem, Combinatorica 4, 247--257 (1984)

\bibitem{wolf} Wolf, J.: Born again group testing: multiaccess communications.     IEEE Trans. Inf. Theory 31, pp. 185--191  (1985)
\bibitem{zrb} Zhu, J., Rivera, K., and Baron, D.:  Noisy pooled pcr for virus testing. Available at 
https://doi.org/10.48550/arXiv.2004.02689 (2020)
\end{thebibliography}
\end{document}